\documentclass[%
 reprint,
showpacs,
 amsmath,amssymb,
 aps,
 pra,
]{revtex4-1}

\usepackage[cmyk]{xcolor}

\usepackage{graphicx} 
\usepackage{dcolumn} 
\usepackage{bm} 

\usepackage[utf8]{inputenc}
\usepackage[T1]{fontenc}
\usepackage{times}
\usepackage{flafter} 
\usepackage{textgreek} 
\usepackage{mathrsfs} 

\newcommand{\ii}{\mathrm{i}}

\newcommand{\itPsi}{\mathit{\Psi}}

\newcommand{\itGamma}{\mathit{\Gamma}}
\newcommand{\itOmega}{\mathit{\Omega}}
\newcommand{\itDelta}{\mathit{\Delta}}

\newcommand{\derivative}{\partial}

\newcommand*\diff{\mathop{}\!\mathrm{d}}

\makeindex

\begin{document}

\title{Effects of inner electrons on atomic strong-field ionization dynamics}

\author{J.\ Rapp}
\author{D.\ Bauer}%
\affiliation{%
 Institut für Physik, Universität Rostock, 18051 Rostock, Germany
}%


\date{\today}

\begin{abstract}
The influence of inner electrons on the ionization dynamics in strong laser fields is investigated in a wavelength regime where the inner electron {\em dynamics} is usually assumed to be negligible. The role of inner electrons is of particular interest for the application of frozen-core approximations and pseudopotentials in time-dependent density functional theory (TDDFT) and the single-active-electron (SAE) approximation in strong-field laser physics. Results of TDDFT and SAE calculations are compared with exact ones obtained by the numerical {\em ab initio} solution of the three-electron time-dependent Schrödinger equation for a lithium model atom. It is found that dynamical anti-screening, i.e., a particular form of dynamical core polarization, may substantially alter the ionization rate in the single-photon regime.  Requirements for the validity of the approximations in the single and multiphoton ionization domain are identified.  
\end{abstract}

\pacs{
  32.80.Fb 
, 32.80.Rm 
, 31.15.ee 
, 31.15.es 
}%

\maketitle

\section{Introduction}
Density functional theory (DFT) simulations (see, e.g., \cite{dft-parr-yang,dft-old,dft-engel-dreizler}) have become popular tools for electronic structure calculations. Compared to the exact solution of the many-body Schrödinger equation, discrepancies in the electron density obtained from DFT-based Kohn-Sham (KS) \cite{KS} schemes  are, by construction, caused by the unavoidable  approximation to the generally unknown exact exchange-correlation (XC) potential. However, in practice it is common to apply additional approximations, most notably ``pseudopotentials'' or the ``frozen-core approximation'' (see, e.g., \cite{pseudo-szasz,pseudopotentials}) in order to avoid the numerical effort of treating tightly-bound inner electrons. The justification for this neglect is that core electrons  do not take directly part in, e.g.,  the formation of bonds. 

DFT has been extended to systems in time-dependent external potentials. In principle, time-dependent density functional theory (TDDFT) (see, e.g., \cite{tddft-fundamentals,tddft-new})  allows to study many-electron systems such as atoms, molecules, or clusters in strong laser fields, even beyond linear response. It is known, however, that switching from DFT to TDDFT makes the unknown exact XC functional even more inaccessible because of memory effects and the initial-state dependence it should contain \cite{ullrich:234108,hessler:72,PhysRevB.73.075413,PhysRevLett.89.023002-Maitra-Burke,tddft-new} but all practicable approximations to it do not. 

In this paper, we study inner-electron dynamics induced by time-dependent external fields. In contrast to typical applications of DFT concerning the ground state electronic structure of the system at hand, even the lowest KS orbitals may undergo a significant modification if the system is subjected to a strong external laser field in TDDFT beyond linear response. Evidently, it is invalid to freeze those KS orbitals which directly contribute to, e.g.,  the outgoing electron density of an atom being ionized. If, on the other hand, solely the KS  valence orbital dominates ionization, it is an eligible question if the essential dynamics can be reproduced by a frozen-core, pseudo, or single-active-electron (SAE) potential.  In fact, the SAE approximation is ubiquitous in the strong-field ionization community (see, e.g., \cite{0034-4885-60-4-001,0953-4075-39-14-R01,QUA:QUA560400839,PhysRevA.77.063403}). Only recently it has been recognized that in multi-electron molecules it is often not permissible to consider only the highest occupied molecular orbital in strong-field processes \cite{nature-olga,Boguslavskiy16032012}.   

The  question we address in this paper is whether core electrons in atoms can be considered ``frozen'' or not in the interaction with {\em long} wavelength radiation. Here,  ``long'' means that the photon energy $\hbar\omega$ should be small compared to the energy by which the core electrons are bound. Given the energy level spacings of the Li atom,  we thus need to consider the multiphoton regime, and the single-photon regime up to photon energies well below values where the core electrons are accessed ``directly'' by the applied laser field.

 We employ  a model Li atom, for which we are able to numerically solve the time-dependent Schr\"odinger equation (TDSE) {\em ab initio}. Lithium is the simplest element with core electrons in the ground state configuration and thus serves as a perfect testing ground. However, strong-laser driven Li in full dimensionality is well beyond nowadays computational capabilities. Even with the spatial degrees of freedom restricted to one dimension (i.e., the laser polarization direction) per electron, the numerical demand is enormous  for strong laser fields. We have been able to speed up the calculations by employing properties of the time-dependent, spatial three-body wavefunction in the ionization regime considered and by optimizing the TDSE solver for graphics processing units. 

The paper is organized as follows. In Sec.~\ref{chapter-model} the Li model system is described. In Sec.~\ref{chapter-methods} the methods and approximations used in this work are introduced. Results are presented in Sec.~\ref{sec-results},  a conclusion and outlook are given in Sec.~\ref{sec-concl}. Remarks on numerical details are attached as an Appendix.

Atomic units (a.u.) are used throughout.

\section{One-dimensional lithium model}\label{chapter-model}
The Li atom is the simplest atom with ``inner'' and ``outer'' electrons in the ground state configuration.  The reduction to one dimension per electron is required for the exact numerical treatment, as the computational effort grows exponentially with both particle number and dimension. One-dimensional atom models have been successfully used in the case of helium for more than 20 years~\cite{1dhelium} and more recently for Li as well~\cite{1dlithium}.

Applying the dipole approximation, the Hamiltonian in length gauge  reads
\begin{equation}%
 H(t)=\sum_i \left(T^{(i)}+V^{(i)}+H_\mathrm{L}^{(i)}(t)+\frac{1}{2}\sum_{j\neq i}W^{(ij)}\right) \label{lithium-ham}
\end{equation}%
with indices $i,j\in\left\{1,2,3\right\}$, the kinetic energy operator
\begin{equation}
 T^{(i)}=\frac{1}{2}\left(p^{(i)}\right)^2,
\end{equation}
the core potential 
\begin{equation}
 V^{(i)}=-Z\left[\left(x^{(i)}\right)^2+\varepsilon^2\right]^{-1/2}, \quad Z=3,
\end{equation}
the coupling to the laser field
\begin{equation}
 H_\mathrm{L}^{(i)}(t)=\mathcal{E}(t)x^{(i)}, \quad \mathcal{E}(t)=-\hat{\mathcal{E}}\sin\omega t,
\end{equation}
and the electron-electron interaction operator $W^{(ij)}$ 
\begin{equation}
 W^{(ij)}=\left[\left(x^{(i)}-x^{(j)}\right)^2+\varepsilon^2\right]^{-1/2}.
\end{equation}
The smoothing parameter $\varepsilon=0.5034$ is tuned such that the total energy of the ``real,'' three-dimensional Li atom is reproduced.

\section{Methods and approximations}\label{chapter-methods}
In this Section, we introduce the three methods used (TDSE, Floquet, and TDDFT) and the various approximations (frozen-core, pseudopotentials, and SAE). Particular emphasis is put on the structure of the three-electron wavefunction, which can be decomposed into a sum of three terms, each factorizing in a spin and a spatial part.

\subsection{Time-dependent Schrödinger equation}\label{sec-tdse}
The TDSE
\begin{equation}
 \ii\derivative_t|\itPsi(t)\rangle = H(t)|\itPsi(t)\rangle \label{tdse-orig}
\end{equation}
is the fundamental equation describing the non-relativistic time-evolution of a many-particle quantum state $|\itPsi(t)\rangle$. Due to the unavailability of an analytical solution for the Hamiltonian \eqref{lithium-ham} we solve the TDSE numerically. In that way exact benchmark results are obtained to which results from approximate approaches will be compared.

\subsubsection{Three-electron state $|\itPsi(t)\rangle$}
Let us expand the state $|\itPsi(t)\rangle$ in orthonormal single-particle states 
\begin{equation}
 \{|n\rangle\}_{n\in\mathbb{N}},
  \qquad
 |n\rangle=|\phi_n\rangle\otimes|\chi_n\rangle,
\end{equation}
where $|\phi_n\rangle$ and  $|\chi_n\rangle$ are spatial and spin components, respectively. Suppressing the time-argument, the expansion for three particles reads
\begin{eqnarray}
 |\itPsi\rangle &=& \sum_n\left[\left(\sum_k a_{kn} |k\rangle^{(1)}\right)\right.\nonumber\\
  &&\left.\otimes\left(\sum_l b_{ln} |l\rangle^{(2)}\right)\otimes\left(\sum_m c_{mn} |m\rangle^{(3)}\right)\right] \label{state-correlation}\\
  &=& |ABC\rangle \label{state-abbreviation}.
\end{eqnarray}
For brevity, the $\otimes$-sign  denoting the tensor product will be omitted from now on. The shorthand notation \eqref{state-abbreviation} allows to concisely formulate the correct exchange antisymmetry in the case of fermions
\begin{eqnarray}
 |ABC\rangle &=& -|ACB\rangle = |CAB\rangle\nonumber\\*
   &=& -|BAC\rangle = |BCA\rangle = -|CBA\rangle \label{state-antisymmetry}.
\end{eqnarray}
The antisymmetry \eqref{state-antisymmetry} can be enforced on a general three-particle state $ |A^\prime B^\prime C^\prime\rangle$ by the antisymmetrization operator $\mathcal{A}$,
\begin{equation}
 |\itPsi\rangle = |ABC\rangle = \mathcal{N}^\prime\mathcal{A} |A^\prime B^\prime C^\prime\rangle,
\end{equation}
where the normalization factor $\mathcal{N}^\prime$ has to be chosen such that $\langle\itPsi|\itPsi\rangle=1$,
and 
\begin{eqnarray*}
 \mathcal{A} |A'B'C'\rangle &=& \frac{1}{3!}\big(|A'B'C'\rangle - |A'C'B'\rangle + |C'A'B'\rangle\nonumber\\*
  &&- |B'A'C'\rangle + |B'C'A'\rangle - |C'B'A'\rangle\big).
\end{eqnarray*}

Introducing the abbreviation $|\boldsymbol{x\sigma}\rangle=|x_1\rangle|\sigma_1\rangle|x_2\rangle|\sigma_2\rangle|x_3\rangle|\sigma_3\rangle$, the expansion of $|\itPsi\rangle$ in position-spin space reads
\begin{equation}
 |\itPsi\rangle = \sum_{\sigma_1\sigma_2\sigma_3}\iiint\diff x_1\diff x_2\diff x_3\,|\boldsymbol{x\sigma}\rangle\langle\boldsymbol{x\sigma}|\mathcal{N}^\prime\mathcal{A}|A^\prime B^\prime C^\prime\rangle \label{wf-expand}.
\end{equation}
The configuration is chosen such that the total spin is $S=1/2$ and $M_S=+1/2$ at all times. This can be justified by the fact that interaction Hamiltonians  that could induce spin-flips are not considered in this paper. The primed state $|A^\prime B^\prime C^\prime \rangle$ can be chosen to have separable spin components, e.g., the corresponding expansion coefficients $a_{kn}^\prime$ are only non-vanishing for spin-down while the other two coefficients always result in spin up. As a consequence, the function
\begin{equation}
\itPsi(x_1, x_2, x_3)=\sum_{\sigma_1\sigma_2\sigma_3}|\boldsymbol{\sigma}\rangle\langle\boldsymbol{x\sigma}|\mathcal{N}^\prime\mathcal{A}|A^\prime B^\prime C^\prime\rangle
\end{equation}
can be written as
\begin{align}
 \itPsi(x_1, x_2, x_3) &= \frac{\mathcal{N}^\prime}{3!}\Big[
    \left|\downarrow\uparrow\uparrow\right\rangle\big(\phi(x_1, x_2, x_3)-\phi(x_1, x_3, x_2)\big)  \nonumber\\*
    &\quad+ \left|\uparrow\downarrow\uparrow\right\rangle\big(\phi(x_2, x_3, x_1)-\phi(x_2, x_1, x_3)\big)  \nonumber\\*
    &\quad+ \left|\uparrow\uparrow\downarrow\right\rangle\big(\phi(x_3, x_1, x_2)-\phi(x_3, x_2, x_1)\big)
  \Big],
\end{align}
with correlated spatial functions $\phi(x_1, x_2, x_3)=\sum_{klmn} a_{kn}^\prime b_{ln}^\prime c_{mn}^\prime  \langle x_1|\phi_k\rangle\langle x_2|\phi_l\rangle\langle x_3|\phi_m\rangle$. Defining 
\begin{equation}
 \phi_{23}(x_1, x_2, x_3)=\mathcal{N}^{\prime\prime}\big[\phi(x_1, x_2, x_3)-\phi(x_1, x_3, x_2)\big],
\end{equation}
which is antisymmetric with respect to the exchange of its second and third argument,
one obtains the compact form
\begin{eqnarray}
 \itPsi(x_1, x_2, x_3) &=& \mathcal{N}\Big[
    \left|\downarrow\uparrow\uparrow\right\rangle\phi_{23}(x_1, x_2, x_3)\nonumber\\*
    &&+ \left|\uparrow\downarrow\uparrow\right\rangle\phi_{23}(x_2, x_3, x_1)\nonumber\\*
    &&+ \left|\uparrow\uparrow\downarrow\right\rangle\phi_{23}(x_3, x_1, x_2)
  \Big], \label{wf-compact}
\end{eqnarray}
where $\mathcal{N}=\tfrac{\mathcal{N}^\prime}{\mathcal{N}^{\prime\prime}3!}$. The full three-electron state $|\itPsi\rangle$ is---at all times---completely determined by $\phi_{23}$,
\begin{equation}
 |\itPsi\rangle = \mathcal{N}\Big[
    1 + \mathcal{P}^{(12)}\mathcal{P}^{(23)} + \mathcal{P}^{(23)}\mathcal{P}^{(12)}
 \Big] \left|\downarrow\uparrow\uparrow\right\rangle|\phi_{23}\rangle. \label{state-final}
\end{equation}
Here, $P^{(ij)}$ is the two-particle permutation operator which exchanges the indices of particles $i$ and $j$,
and
\begin{equation}
 |\phi_{23}\rangle = \iiint\diff x_1\diff x_2\diff x_3 |\boldsymbol{x}\rangle\phi_{23}(x_1, x_2, x_3).
\end{equation}
We assume that $|\phi_{23}\rangle$ is normalized to unity, $\langle\phi_{23}|\phi_{23}\rangle=1$, so that  $\mathcal{N}=\tfrac{1}{\sqrt{3}}$. 


\subsubsection{Spatial TDSE}
Inserting a time-dependent state $|\itPsi(t)\rangle$ of the form \eqref{state-final} into the TDSE \eqref{tdse-orig} yields
\begin{equation}
 \Big[
    1 + \mathcal{P}^{(12)}\mathcal{P}^{(23)} + \mathcal{P}^{(23)}\mathcal{P}^{(12)}
 \Big]
  \Big[\ii\derivative_t-H(t)\Big]\left|\downarrow\uparrow\uparrow\right\rangle|\phi_{23}(t)\rangle=0 \label{tdse-intermediate}
\end{equation}
because both $\derivative_t$ and  $H(t)$ commute with any two-particle permutation operator $\mathcal{P}^{(ij)}$. 

Assuming a spin-diagonal Hamiltonian, multiplication of \eqref{tdse-intermediate} from the left by, e.g., $\left\langle\downarrow\uparrow\uparrow\right|$, yields a TDSE for the evolution of $|\phi_{23}(t)\rangle$ in time,
\begin{equation}
 \ii\derivative_t|\phi_{23}(t)\rangle = \left\langle\downarrow\uparrow\uparrow\right|H(t)\left|\downarrow\uparrow\uparrow\right\rangle|\phi_{23}(t)\rangle \label{tdse-like-equation}.
\end{equation}
Although the Hamiltonian \eqref{lithium-ham} does not act on spin components at all, the TDSE \eqref{tdse-like-equation} still holds for spin-diagonal Hamiltonians. This will be utilized in Sec.~\ref{sec-gedankenexperiment}.  The TDSE \eqref{tdse-like-equation} is the one that is actually solved numerically in position space on a discretized $x_1x_2x_3$-grid. More details about the numerical solution are described in the Appendix.

\subsubsection{Observables for both spin projections}
Although electrons are indistinguishable, the partial wavefunction $\phi_{23}$ allows to extract information about inner and outer electrons separately. Given a one-particle operator $a^{(i)}$ acting on the spatial component only, one can construct a spin-spatial operator $a_\sigma$ of the form
\begin{equation}
 a_\sigma=\sum_i\left|\sigma\right\rangle^{(i)}\left\langle\sigma\right|^{(i)}a^{(i)}, \label{operator-with-spin-part}
\end{equation}
where 
\begin{align*}
 \left|\sigma\right\rangle^{(1)}\left\langle\sigma\right|^{(1)} &= \sum_{\sigma_1\sigma_2} \left|\sigma\right\rangle\left|\sigma_1\right\rangle\left|\sigma_2\right\rangle\left\langle\sigma\right|\left\langle\sigma_1\right|\left\langle\sigma_2\right|, \\
 \left|\sigma\right\rangle^{(2)}\left\langle\sigma\right|^{(2)} &= \sum_{\sigma_1\sigma_2} \left|\sigma_1\right\rangle\left|\sigma\right\rangle\left|\sigma_2\right\rangle\left\langle\sigma_1\right|\left\langle\sigma\right|\left\langle\sigma_2\right|, \\
 \left|\sigma\right\rangle^{(3)}\left\langle\sigma\right|^{(3)} &= \sum_{\sigma_1\sigma_2} \left|\sigma_1\right\rangle\left|\sigma_2\right\rangle\left|\sigma\right\rangle\left\langle\sigma_1\right|\left\langle\sigma_2\right|\left\langle\sigma\right|.
\end{align*}
The choice of either $\left|\sigma\right\rangle=\left|\downarrow\right\rangle$ or $\left|\sigma\right\rangle=\left|\uparrow\right\rangle$ then yields an operator for calculating observables for the {\em single spin-down inner electron} on the one hand and for the two spin-up electrons on the other hand, respectively. The latter will be referred to as \emph{inner-outer spin component} \footnote{Had we chosen the Li configuration $S=1/2$ and $M_S=-1/2$ instead of  $M_S=+1/2$, the role of single-particle spin-up and spin-down components would be reversed, of course.}.

\subsubsection{Ionization}\label{sec-channels}
\begin{figure}[htbp]%
 \includegraphics[width=\columnwidth]{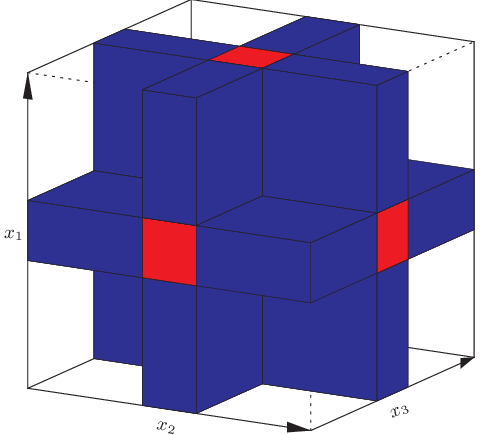}%
 \caption{(Color online) Schematic view of a cubic simulation box around the nucleus in the center. Different colors indicate those regions where zero (neutral Li), one (Li$^+$), two (Li$^{2+}$), or all three electrons (Li$^{3+}$) are located at positions far from the nucleus, respectively.}%
 \label{fig-box}%
\end{figure}%
The ionization probability is chosen as the primary observable for our investigations because it is well-defined and comparable among all considered methods. In the TDSE simulation, position space $(x_1, x_2, x_3)$ is divided into four regions, differing by the number of electrons which are located far away from the nucleus (see Fig.~\ref{fig-box}). The respective ionization regions are
(i) no ionization: single small cube around the nucleus,
(ii) single ionization: six channels pointing to the surface centers of the simulation box,
(iii) double ionization: twelve cuboids, four lying in each of the three central plains,
(iv) triple ionization: eight cubes in the corners of the simulation box.

The laser parameters considered throughout this paper are such that multiple ionization is negligible. Thus, the (single-) ionization probability $p(t)$ reduces to $p(t)=1-N(t)$ where $N(t)$ is the norm inside the cube around the nucleus (representing neutral Li).

\subsection{Floquet method}\label{sec-floquet}
The whole purpose of applying the Floquet approach in this work is the determination of resonances, taking the AC Stark effect into account.

In general, the Floquet method (see, e.g., \cite{joachainpotvl,Chu20041})  allows to study time-dependent problems without an explicit time-propagation. This is possible if the Hamiltonian $H(t)$ is periodic in time, i.e.
\begin{equation}
 H(t)=H_0+H_\mathrm{L}(t),
  \qquad
 H_\mathrm{L}(t+T)=H_\mathrm{L}(t),
\end{equation}
because the so-called Floquet theorem then allows to obtain a time-independent eigenvalue equation for the field-dressed states and the corresponding quasienergies.  Quasienergy spectra are useful for predicting resonance enhancements in the ionization rate as a function of the laser frequency with the AC Stark effect automatically included.
The Floquet method will thus  be used to follow the quasienergies $\epsilon_m^{(n)}$ for varying photon energies $\omega$ and a fixed electric field amplitude $\hat{\mathcal{E}}$. Here, the index $m\in\mathbb{N}$  refers to the (unperturbed) atomic energy level, the index $n\in\mathbb{Z}$ to the ``Floquet block'' (note that  $\epsilon_m^{(n)}+\omega=\epsilon_m^{(n+1)}$).

\subsection{Time-dependent density functional theory}\label{sec-tddft}
DFT \cite{dft-parr-yang,dft-old,dft-engel-dreizler} and its time-dependent extension TDDFT \cite{tddft-fundamentals,tddft-new} are approaches to overcome the exponential scaling of the numerical effort with the number of particles. They are based on the fact that all information about the system is included in the (time-dependent) single-particle density $n(x)$, as proven by the Hohenberg-Kohn theorem~\cite{hohenbergKohn} and its time-dependent analog, the Runge-Gross theorem~\cite{rungeGross}. (TD)DFT calculations are performed in practice within a KS scheme, i.e., the single-particle density is reproduced with the help of a fictitious, much easier-to-solve noninteracting system evolving in an effective KS potential.

In this work, the spin-polarized Li-system is studied.  We therefore consider the spin-densities $n_\sigma(x)$, $\sigma\in\{\downarrow,\uparrow\}$. The exchange-only local-spin-density approximation (LDA) is employed for the XC functional. The exchange functional for the three-dimensional electron gas is considered because the one-dimensional model introduced in Sec.~\ref{chapter-model} is meant to mimic a three-dimensional three-electron  atom in a linearly polarized laser field rather than a true one-dimensional system.

The time-dependent KS equation reads (spatial arguments suppressed)
\begin{equation}
\ii \partial_t \varphi_i(t)  = H^{(\sigma_i)}_\mathrm{KS}(t) \varphi_i(t)
\end{equation}
with the KS Hamiltonian
\begin{eqnarray}
H^{(\sigma_i)}_\mathrm{KS}(t) &=& - \frac{1}{2} \frac{\partial^2}{\partial x^2} + v(t)  \\
&&  + v^{(\mathrm{H})}[n(t)] + v^{(\mathrm{XC})}[n_{\sigma_i}(t)]\nonumber
\end{eqnarray}
and
$ n_{\sigma}(t) = \sum_{i} n_i(t) \,\delta_{\sigma_i\sigma}$, $n_i(t)=|\varphi_i(t)|^2$, $n(t)=\sum_\sigma n_{\sigma}(t)$.  The same external potential $v(t)$ as in the many-body TDSE (i.e., binding potential plus laser in our case) appears here, $ v^{(\mathrm{H})}$ is the Hartree potential, and $ v^{(\mathrm{XC})}$ is the XC potential (to be approximated).

A known problem of LDA is the wrong asymptotic behavior of the KS potential $v+v^{(\mathrm{H})}+v^{(\mathrm{XC})}$. Each KS orbital in an, e.g., unperturbed, neutral atom   should experience a potential $-{1}/{r}$ far away from the nucleus, representing one unscreened nuclear charge. This correct behavior can be enforced by the so-called Perdew-Zunger (PZ) self-interaction correction (SIC) \cite{pzSic}. PZ SIC corrects the Hartree and XC potentials for each orbital $i$ by subtracting the potentials evaluated for the ``own'' single-particle density $n_i$. In the case of PZ-corrected LDA, the final corrected XC potential is calculated as (time and space arguments suppressed)
\begin{eqnarray}
 v_i^{(\text{LDA+PZ})}&=&v^{(\mathrm{LDA})}[n_{\sigma_i}]-v^{(\mathrm{H})}[n_i] -v^{(\mathrm{LDA})}[n_i].\quad
\end{eqnarray}
It is easy to see that this leads to the correct SIC in the single-electron limit. However, due to the nonlinearity of the KS potential, i.e.,  $v^{(\mathrm{XC})}[n]-v^{(\mathrm{XC})}[n_i]\neq v^{(\mathrm{XC})}[n-n_i]$, the self-interaction is not completely removed by PZ-SIC in general.
The PZ-corrected KS Hamiltonian is orbital-dependent, i.e., it is in general different for each orbital $i$ (i.e., not only different for orbitals with different $\sigma_i$, as in ``conventional'' spin-DFT). An unpleasant consequence of this orbital-dependence of the KS Hamiltonian is the non-orthogonality of the PZ-SIC KS orbitals (although in practice they are usually very close to orthogonal). Positive consequences of the PZ-SIC are that, besides the correct asymptotic behavior of potentials and densities, the values for the total energy, and ionization energies  (or electron affinities of negative ions) typically improve.

\subsection{Frozen-core approximation}
All-electron (TD)DFT calculations often become numerically too demanding. Hence, it is common to apply additional approximations in order to reduce the numerical effort further. One may employ the fact that chemical bonds and reactions are governed by valence electrons so that the relaxation of electronic core shells may be considered negligible. Consequently, tightly localized KS orbitals may be  self-consistently determined for the initial state configuration but ``frozen'' during the actual TDDFT time-propagation.  This approach will be referred to as ``frozen-core approximation'' (FCA).
One of the issues in this work is the validity of the FCA for atoms interacting with a strong laser field, i.e., in  TDDFT calculations beyond linear response.

\subsection{Pseudopotentials}
Consider a  one-dimensional (e.g., radial) KS orbital, which is orthogonal to $n$ other mutually orthogonal orbitals. If these other orbitals are constructed with the minimum number of nodes the considered orbital must have at least $n$ nodes. A numerical grid thus requires a fine spatial resolution in order to resolve all orbital nodes, including the behavior in-between where the second derivative can reach high absolute values.

A popular tool which aims at circumventing the numerical demand caused by orbitals with many, densely-distributed nodes are ``pseudopotentials'' (see, e.g., \cite{pseudo-szasz,pseudopotentials}). After applying the FCA, an artificial potential is constructed such that those orbitals which are not frozen yield the same single-particle density outside  a certain cutoff radius $r_\mathrm{c}$ as in the full calculation but have less nodes within  $[0,r_\mathrm{c}]$. Moreover, the pseudopotential is ``designed'' to reproduce the KS energies of the unfrozen orbitals (and possibly also those of the unpopulated, excited states). As in the motivation of the FCA, the argument for using pseudopotentials in chemistry is that only valence electron densities are important for the questions of interest such as molecular binding properties and chemical reactions. In other words, only the electron density far away from the nuclei is important.

In the hierarchy of approximations, pseudopotentials reside below the FCA. Hence we do not test particular pseudopotentials in this work. If the FCA fails, pseudopotentials will fail too (unless there is a lucky cancellation of errors caused by the removal of the nodes for $r<r_\mathrm{c}$).

\subsection{Single-active-electron approximation}\label{sec:sae}
In the single-active-electron (SAE) picture it is assumed that the electron under investigation can be described as a single particle moving in an effective, external potential. It thus may be also viewed as an FCA. DFT provides one option to approximate this effective, external potential: one performs an all-electron DFT calculation for the desired initial electron configuration (usually the ground state) and subsequently ``freezes'' all KS orbitals but the one for the SAE of interest for the real-time propagation \footnote{Rigorously speaking, the fictitious KS particles should not be viewed as ``real'' electrons. However, it seems not unreasonable to take the valence KS orbital as the starting point for a SAE calculation.}. In that way, a dependence on the XC potential is introduced only indirectly through the initial state. In the strong-field community, often simple analytical expressions for screened Coulomb potentials with adjustable screening parameters are used \cite{PhysRevA.60.1341,PhysRevA.54.3261}. 

The question arises which state the SAE should populate in the effective potential. Common choices are (i) the ground state (corresponding to a pseudopotential approach with one valence electron) or (ii) some excited state (corresponding to pure FCA). 
The latter choice is more relevant for applications to intense laser-atom interaction, as the orbital symmetry of the initial state of the valence electron is important and can be measured in ionization experiments \cite{Meckel13062008}. We will therefore concentrate on this case.

\section{Results} \label{sec-results}
The results in this work are organized as follows.  In Sec.~\ref{sec-results-groundstate} the lowest-lying stationary states of the model Li atom introduced  in Sec.~\ref{chapter-model} are determined. Exact results for the ionization rate are considered in Sec.~\ref{sec-results-ionization} in order to identify different mechanisms behind the ionization process for different regimes of laser parameters. In Sec.~\ref{sec-gedankenexperiment}, a \emph{ge\-dan\-ken\-ex\-peri\-ment} is performed, revealing the mechanisms by which seemingly passive ``inner'' electrons can influence the ionization probability of the ``outer'' electron. In Sec.~\ref{sec-results-tddft} the exact results are compared with TDDFT in various approximations.


\subsection{Stationary states}\label{sec-results-groundstate}
The highly optimized TDSE solver for propagating the  full three-particle wavefunction (details in the Appendix) is used in the imaginary-time mode for calculating the unperturbed eigenstates of the model Li.
The ground state energy is $E_0=-7.4782$.  As we are investigating single ionization in the present work, it is useful to determine the single-ionization continuum threshold. This can be done by comparing with the ground state of Li$^+$ or by following the Rydberg series of singly-excited states $E_m$ of the neutral Li towards $E_\infty$, both leading to  an ionization potential $E_\mathrm{ip}=E_\infty-E_0\simeq 0.375$.  Table~\ref{table-groundstate} lists the energies for the lowest eight excited states.

\begin{table}[htbp]
  \caption{Energies $E_m$ of the energetically lowest eigenstates $m=0,1,2, \ldots 8$ of  the Li model atom.}
  \begin{tabular}{ p{2cm}  p{2cm} }\hline\hline
    $m$ & \quad $E_m$ \\
    \hline
    $0$ & $-7.4782$\\
    $1$ & $-7.2838$\\
    $2$ & $-7.2020$\\
    $3$ & $-7.1657$\\
    $4$ & $-7.1431$\\
    $5$ & $-7.1306$\\
    $6$ & $-7.1213$\\
    $7$ & $-7.1156$\\
    $8$ & $-7.1110$\\\hline
  \end{tabular}
  \label{table-groundstate}
\end{table}

As an example, Fig.~\ref{fig-plot3d} in the Appendix shows a cut  at $x_1=0$ of $|\phi_{23}(x_1, x_2, x_3)|^2$ for the seventh excited state. The single spin-down component is oriented spatially along $x_1$  in the partial wavefunction $\phi_{23}$. As the spin-down component necessarily belongs to an inner electron,  the extent of the probability density $|\phi_{23}|^2$ is small in $x_1$-direction. Hence the cut at  $x_1=0$. We further observe  the antisymmetry plane $x_2=x_3$ and the preference of spatial  regions where no more than one electron is located at a position comparatively far from the nucleus.


\subsection{Ionization in different laser regimes}\label{sec-results-ionization}
The ionization rate $\itGamma_\omega$ for a certain photon energy $\omega$ was determined by fitting $N(t)$ to  $p_\omega(t)\simeq 1-\exp(-\itGamma_\omega t)$ (see the definition of $p(t)$ in Sec.~\ref{sec-channels}) during the  flat-top part of a trapezoidal laser pulse. 
Given a maximum simulation time $\tau\geq t\geq0$, the electric field amplitude $\hat{\mathcal{E}}$ during the flat-top part of the pulse is chosen high enough to make the relevant ionization timescales for a ``numerically measurable'' ionization yield shorter than $\tau$. Besides the inverse ionization rate $\itGamma^{-1}_\omega$  also the inverse $n$-photon Rabi frequency $\itOmega_{\mathrm{R},n}^{-1}$ matters here.
On the 
other hand, the ponderomotive energy $U_\mathrm{p}=\tfrac{\hat{\mathcal{E}^2}}{4\omega^2}$ should be smaller than both the ionization potential $E_\mathrm{ip}$ and the photon energy $\omega$. Otherwise, (strong field) effects such as AC Stark shifts, above-threshold ionization, and stabilization \cite{bauerbuch} could influence the ionization dynamics dramatically \footnote{Note, however, that all methods used in this work can be applied to the nonperturbative regime as well. We restrict ourselves to rather small field strengths here for purely methodical reasons.}. The electric field amplitude was therefore  set to $\hat{\mathcal{E}}=0.05$.

It is obvious that FCAs fail for photon energies $\omega\gg E_\mathrm{ip}$ high enough to produce core holes. Too low laser frequencies are numerically too demanding. Hence, the frequencies considered in this work are restricted to $\omega\in[0.1,1.0]$. Within this regime we encounter single-photon ionization for $\omega\geq E_\mathrm{ip}$ and multiphoton ionization for $\omega<E_\mathrm{ip}$.

\subsubsection{Ionization rates}\label{sec-ionization-rates}
\begin{figure}[htbp]%
   \includegraphics[width=\columnwidth]{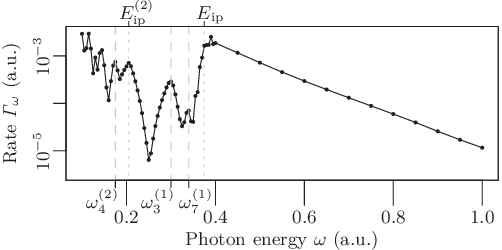}%
   \caption{Logarithmic plot of the ionization rate $\itGamma_\omega$ of the Li model atom vs the laser frequency $\omega$ for   $\hat{\mathcal{E}}=0.05$, as obtained from the {\em ab initio} solution of the TDSE. Ionization thresholds for one and two photons are given by the dotted, vertical lines. $n$-photon resonances with the $m^\text{th}$ excited state are denoted by $\omega_m^{(n)}$ (dashed, vertical lines). The regime of multiphoton ionization $\omega<E_\mathrm{ip}$ is dominated by resonances at $\omega_m^{(n)}$ and $E_\mathrm{ip}^{(2)}$ whereas the photoionization probability decreases exponentially (i.e., linearly on the logarithmic scale) for $\omega>E_\mathrm{ip}$, starting at its maximum for $\omega\simeq E_\mathrm{ip}$.  }%
   \label{fig-rates-exact}%
\end{figure}

Ionization rates $\itGamma_\omega$ obtained by the exact solution of the TDSE  are shown in Fig.~\ref{fig-rates-exact}. As expected, one can qualitatively distinguish between the single-photon and multiphoton  ionization regime.

\paragraph{Single-photon ionization.}
 If the photon energy increases beyond $\omega=E_\mathrm{ip}$, the ionization rate drops exponentially.  Note that this behavior can only partially be explained by the decreasing number of photons per time and area for the fixed laser intensity in the simulation. In a simple picture, the ionization rate $\itGamma_\omega$ should be the product of the photoionization cross section $\sigma=\sigma(\omega)$ and the photon impact rate per area $\itGamma_\mathrm{photon}/A$. Hence one would expect $\itGamma_\omega=\sigma\itGamma_\mathrm{photon}/A=\sigma I/\omega$, where $I$ denotes the (in our case constant) laser intensity. Instead, the almost linearly decreasing  slope in the logarithmic plot of $\itGamma_\omega$  in Fig.~\ref{fig-rates-exact} shows that $\itGamma_\omega$ is {\em not} $\sim \omega^{-1}$. Hence there must be an exponential dependence in $\sigma(\omega)$.

\paragraph{Multiphoton ionization.}
Peaks in the  ionization rate $\itGamma_\omega$ for $\omega< E_\mathrm{ip}$ can be categorized into two groups.
If the energy of $n$ photons is just sufficient to free the outer electron, the ionization probability is particularly high. Consequently, one finds a peak just above the two-photon ionization threshold $E_\mathrm{ip}^{(2)}\simeq {E_\mathrm{ip}}/{2}$. However, the AC Stark shift increases for smaller laser frequencies $\omega$. Thus, it becomes more important to consider field-dressed states in order to predict $n$-photon ionization thresholds for $n> 2$.

Another mechanism that leads to peaks in the multiphoton regime is excited-state-assisted ionization where  (a) the energy of $n$ photons matches the energy gap $E_m-E_0$ between the ground state and the $m^\text{th}$ excited state, (b) the $n$-photon transition between the ground state and the $m^\text{th}$ excited state is allowed, and (c) the binding energy $E_\infty-E_m$ of the $m^\text{th}$ excited state is smaller than the photon energy so that the absorption of one additional photon leads to ionization.
This scenario may be viewed as $n$-photon Rabi oscillations, accompanied by ionization.  Laser-dressed states have to be considered in order to precisely predict the peak positions $\omega_m^{(n)}$, especially for small laser frequencies.

\subsubsection{Time-dependent ionization probability}\label{sec-ionization-probabilities}
\begin{figure}[htbp]%
   \includegraphics[width=\columnwidth]{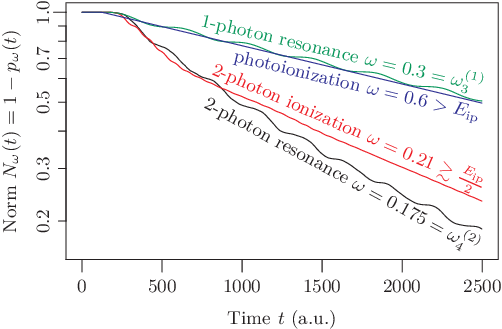}%
   \caption{(Color online) Comparison of time-dependent ionization probabilities $p_\omega(t)$ for photon energies $\omega\in\{0.175, 0.21, 0.3, 0.6\}$ and a fixed electric field amplitude $\hat{\mathcal{E}}=0.05$, ramped over five cycles.  In the case of resonances, i.e., excited-state-assisted ionization, the slope changes periodically with the Rabi frequency. In  one- and two-photon ionization without excited-state assistance, the only relevant timescale is given by $\itGamma_\omega^{-1}$, leading to a  straight slope in the logarithmic plot (disregarding the transient behavior at small times $t$ caused by laser ramping).}%
   \label{fig-norms2}%
\end{figure}

 Ionization just above any $n$-photon ionization threshold should depend solely on a single timescale given by the rate $\itGamma_\omega$. Instead, in the case of excited-state-assisted ionization the  ionization probability vs time 
 should be modulated on the timescale of (multiphoton) Rabi floppings. This is indeed the case, as is shown in  Fig.~\ref{fig-norms2} where the inverse ionization probability $1-p_\omega(t)$ is plotted vs $t$ for four values of the photon energy $\omega$.

\subsubsection{Position of resonance peaks}\label{sec-floquet-results}
\begin{figure*}[htbp]%
   \includegraphics[width=\textwidth]{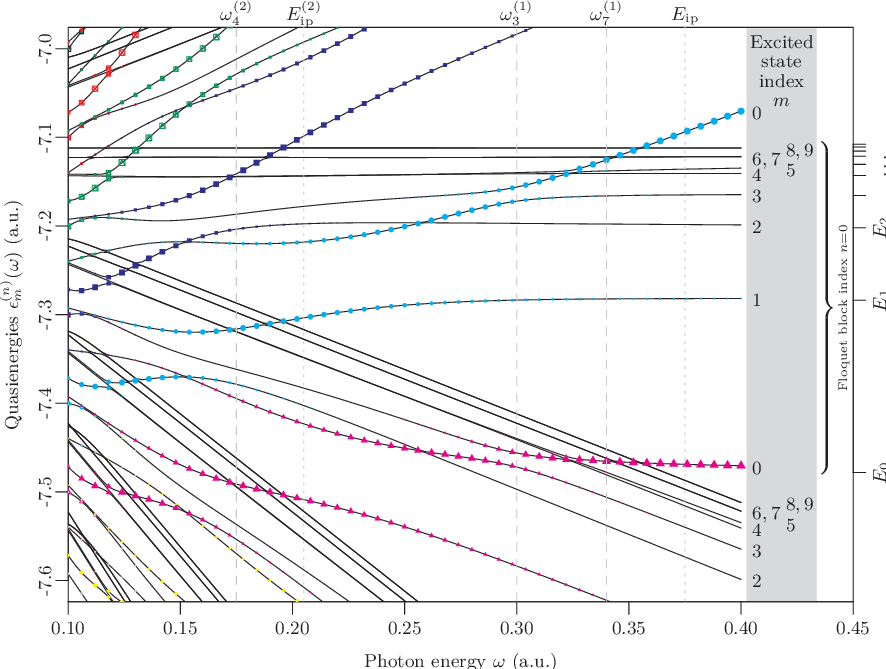}%
   \caption{(Color online) Quasienergies $\epsilon_m^{(n)}(\omega)$ of Floquet states vs photon energy $\omega$ for $\hat{\mathcal{E}}=0.05$. Indices denote the state $m$ and the Floquet block $n$. Those Floquet states with a sizable projection on the atomic ground state $m=0$ are marked by symbols (shaped and colored differently for each Floquet block). Photon energies $\omega$ leading to a peak in the ionization rate $\itGamma_\omega$ are indicated by gray vertical lines, as  in Fig.~\ref{fig-rates-exact}. The energies $E_m$ with $m=0,1,2$ on the right hand side denote unperturbed atomic energy levels.}%
   \label{fig-floquet-follow}%
\end{figure*}

In order to quantitatively predict the position of the peaks in the  ionization rate $\itGamma_\omega$ it is required to consider the AC Stark effect.  The laser parameters used in this work are such that the coupling of the ground state to states with excited inner electrons is negligible. Hence, one can obtain Floquet spectra by considering states below the first ionization threshold only. The results of the Floquet solver are shown in Fig.~\ref{fig-floquet-follow}.

\paragraph{Avoided crossings of the shifted ground state.}
Avoided crossings of the field-dressed ground state are of particular interest here because the system should be described by this state after an appropriate ramping of the laser field in the TDSE solution \cite{PhysRevA.85.023407}.

Following the perturbed ground state $m=0$, $n=0$ from the high-frequency limit to lower frequencies, one  finds avoided crossings with the odd excited states $m\in\{9, 7, 5, 3, 1\}$ of the next lower Floquet block, as expected from the dipole selection rule. The minimum level distances in these one-photon avoided crossings is given by the  Rabi frequency $\itOmega_{\mathrm{R},1}$, which  decreases for increasing $m$ so that those for  $m=9$ and $m=7$ are too close to be resolved. This is expected because $\itOmega_{\mathrm{R},1}$ is proportional to the dipole transition amplitude, which decreases with increasing $m$.

Following in Fig.~\ref{fig-floquet-follow} $m=0$, $n=0$ below the two-photon ionization threshold $E_\mathrm{ip}^{(2)}\simeq 0.2$, two-photon avoided crossings of the ground state with even states $m=8$ (unresolved), $m=6$ (unresolved), $m=4$ (hardly resolved) and $m=2$ (clearly resolved) show up. In the case of $n$-photon crossings for $n\geq 3$, the identification of states becomes cumbersome, as Floquet blocks approach each other and the AC Stark shift increases. 

\paragraph{Prediction of peaks in the ionization rate.}
In Fig.~\ref{fig-floquet-follow}, photon energies with a high  ionization rate are marked by dashed (excited-state-assisted ionization) and dotted ($n$-photon ionization thresholds) vertical lines. For each of these photon energies, the responsible mechanism can be identified by inspecting the behavior of the state $m=0$, $n=0$. The other way around it is not that straightforward. There are avoided crossings at photon energies $\omega_1^{(1)}$ and $\omega_5^{(1)}$ which do not give a significant peak in the  ionization rate $\itGamma_\omega$. However, for most of the avoided crossings one can quantitatively predict a peak position in the ionization rate, which is supporting the mechanisms introduced in Sec.~\ref{sec-results-ionization}.


\subsection{Coupling of inner electrons}\label{sec-gedankenexperiment}
In this Section, a \emph{gedankenexperiment} is performed. Halfway between freezing the inner electrons and taking their dynamics fully into account lies a treatment where only their interaction with the laser is neglected while the electron-electron interaction $W_{ij}$ is included in the simulation. 
 However, switching-off the interaction with the laser for both inner electrons is not possible in the exact TDSE calculation because this would break the exchange symmetry discussed in Sec.~\ref{sec-tdse}. An interaction that does not break the exchange symmetry is a ``spin-selective laser'' coupling $\sum_i H_\mathrm{L}^{(i)}(t)$  with [see Eq.~\eqref{operator-with-spin-part}]
\begin{equation}
 H_\mathrm{L}^{(i)}(t) = \left|\uparrow\right\rangle^{(i)}\left\langle\uparrow\right|^{(i)} \mathcal{E}(t)\,x^{(i)}.
\end{equation}
When solving the TDSE  \eqref{tdse-like-equation} for $\phi_{23}$, this hypothetical laser acts on all electrons except the single inner spin-down electron. In that way we can investigate the role of the (seemingly passive) inner spin-down electron during the ionization process by discriminating its reaction  {\em only} to the spin-up electrons' laser-induced dynamics from its full interaction with {\em both} the laser and the other electrons.

\subsubsection{Ionization rates}\label{sec-ionization}
\begin{figure}[htbp]%
   \includegraphics[width=\columnwidth]{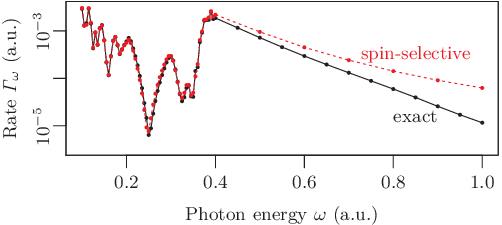}%
   \caption{Ionization rate $\itGamma_\omega$ vs $\omega$ in the case of an artificial, spin-selective laser (dashed) compared to the previous results where all electrons ``see'' the laser (solid).}%
   \label{fig-rates-coupling}%
\end{figure}

Ionization rates for the spin-selective case are compared with the previous full-laser results in Fig.~\ref{fig-rates-coupling}. In the multiphoton ionization regime $\omega<E_\mathrm{ip}$ both curves are in a good agreement. Positions, heights, and widths of the peaks in both cases match quantitatively. However, in the single-photon ionization regime $\omega\geq E_\mathrm{ip}$ significant differences in the ionization rate are observed. Most notably, the  ionization rate for the spin-selective laser is too high and shows a wrong asymptotic behavior with increasing frequency. 
Hence, the interaction of inner electrons with the laser field affects the ionization rate even though all inner electrons stay bound. 

\subsubsection{Dipole expectation values in the laser-driven case}\label{sec-dipoles-laser-driven}
\begin{figure}[htbp]%
\includegraphics[width=\columnwidth]{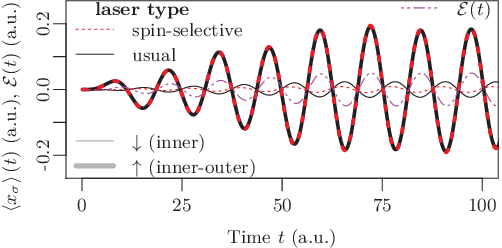}%
   \caption{(Color online) Time-dependent position expectation value of the spin-down component $x_\downarrow(t)$ (inner electron; thin) compared with the position expectation value of the other spin component $x_\uparrow(t)$ (inner and outer electron; bold). The sinusoidal electric field $\mathcal{E}(t)$ with frequency $\omega=0.5$ (thin dashed-dotted, purple) is ramped over five periods and interacts either with all electrons (usual case; solid) or with all electrons except the single spin-down inner electron (artificial spin-selective case; dashed).}%
   \label{fig-dipoles}%
\end{figure}

The excursions of loosely outer and tightly inner bound electrons driven by an oscillating electric field $\mathcal{E}(t)=-\hat{\mathcal{E}}\sin(\omega t)$ are expected to be opposite in phase if the laser period falls in between their respective time scales. In a harmonic binding potential, for example,  the phase depends on the sign of $\omega_0^2-\omega^2$ with $\omega_0$ the eigenfrequency of the harmonic potential.  In the case of a high-frequency driver $\tfrac{\omega_0}{\omega}<1$ the electron is displaced opposite to the driving force $-\mathcal{E}(t)$. On the other hand, a bound electron with a faster timescale than the driver, $\tfrac{\omega_0}{\omega}>1$, is displaced in the direction of the driving force. In terms of inner and outer electrons this means that the position expectation value of an outer electron is more likely to oscillate in phase with the electric field, whereas inner electrons tend to have the opposite phase. Note that ``in phase with the electric field'' means ``opposed to the force'' due to the negative charge of the electron.
 Time-dependent position expectation values for both spin components of the Li model interacting with a ramped sinusoidal laser field are shown in Fig.~\ref{fig-dipoles}. Results were obtained for the case of an ``ordinary'' laser on the one hand and for the case of the artificial spin-selective laser on the other hand.

The amplitude of the single spin-down inner electron is one order of magnitude smaller than the amplitude of the inner-outer spin component. It is thus a good approximation to assign the position expectation value of the inner-outer spin component to the outer electron. As predicted by the harmonic oscillator the loosely bound outer electron oscillates in phase with the electric field whereas the inner spin-down electron  oscillates with the opposite phase.

The comparison of position expectation values for ordinary and spin-selective laser shows that switching-off the laser for the single spin-down inner electron does not affect the oscillation amplitude of the outer electron (see both bold curves on top of each other in Fig.~\ref{fig-dipoles}). In contrast, the amplitude for the single spin-down inner electron itself decreases by a factor of two if the ordinary laser (thin solid black) is replaced by the spin-selective  (thin dashed red).

In the case of the spin-selective laser, the single spin-down inner electron oscillates with the laser frequency $\omega$ although it is not directly interacting with the laser. It only couples indirectly  to the laser field via the electron-electron interaction $W^{(ij)}$. Both spin-up electrons are directly coupled to the laser field and repel the spin-down electron. The latter is therefore slightly displaced in the direction of the electric field by the other inner electron and in the opposite direction by the outer electron. The net result for the quantum mechanical expectation value is an excursion  in the opposite direction (thin  dashed red), but less so as if it were ``seeing'' the laser as well (thin solid black).

As proven by the significantly differing single-photon ionization rates in the {\em gedankenexperiment}, the coupling between inner and outer electrons strongly affects the ionization process despite a seemingly harmless approximation: only the laser interaction of one of the inner electrons is neglected. One can think of the decrease in the ionization rate  in the full simulation as \emph{dynamical anti-screening} of the nuclear charge. This is a particular form of dynamical core polarization,  which could be modeled by adding a polarization potential to the SAE Hamiltonian (see, e.g., \cite{ZhaoYuan2013a,PhysRevLett.111.163001}).  
                                                                                                                    
One might wonder why the ionization rate is so different for the two laser types in Fig.~\ref{fig-rates-coupling}
while the position expectation values for the inner-outer spin-alike electrons in Fig.~\ref{fig-dipoles} are virtually equal. The explanation is that the position   expectation values are dominated by the bound part of the wavefunction while the ionization rate is determined by the (small) escaping (and numerically absorbed) part.                        
                                                                                                                                                                                                                                                                                                                   \begin{figure}[htbp]%
   \includegraphics[width=\columnwidth]{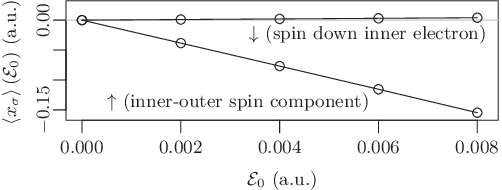}%
   \caption{Dipole expectation values for both spin components of the Li model in the presence of a constant electric field with field strength $\mathcal{E}_0$ (``seen'' by all electrons). The inner-outer spin component is displaced in the direction opposite to the electric field, as expected for negatively charged particles. The displacement of the opposite-spin component (corresponding to the spin-down inner electron here) is in the same direction as the electric field, as a response to the outer electron. }%
   \label{fig-polgame}%
\end{figure}%
\subsubsection{Polarization by a constant electric field}
In the limit $\tfrac{\omega_0}{\omega}\gg1$ a dipole expectation value $x_\sigma(\mathcal{E}_0)=\langle\itPsi(\mathcal{E}_0)|x_\sigma|\itPsi(\mathcal{E}_0)\rangle$ with the sign opposite to $\mathcal{E}_0$ is expected for both spin components if the electrons were non-interacting. However, the electron-electron interaction $W^{{(ij)}}$ modifies this. 

The numerical results  in Fig.~\ref{fig-polgame}   support the picture of inner electrons reacting to the displacement of the outer electron. The expectation value of the single spin-down inner electron $x_\downarrow(\mathcal{E}_0)$ has the same sign as the electric field $\mathcal{E}_0$.


\subsection{Ionization rates obtained with different SAE and TDDFT approximations with and without SIC}\label{sec-results-tddft}
After having obtained insight into the role of inner electrons in ionization and polarization from {\em ab initio} solutions of the TDSE, results from  full TDDFT and frozen-core calculations are presented and interpreted in this Section.  The SAE approximation, as explained in Sec.~\ref{sec:sae}, is also counted as an FCA in which only the KS valence orbital is propagated in the frozen ground state KS potential (plus the potential due to the laser). The SAE results do not suffer from self-interaction introduced during the propagation of orbitals in time. However, a self-interaction error may originate from the  ground state KS  potential. 

For two approximations we considered, the results have been so unreasonable that they are not even shown here but are just mentioned. First, the pure LDA TDDFT approach completely fails in generating a reasonable behavior of ionization probability vs time in the multiphoton regime $\omega<E_\mathrm{ip}$ so that a rate  $p_\omega(t)\simeq 1-\exp(-\itGamma_\omega t)$ could not be extracted. Second, the TDDFT approach using PZ SIC leads to resonances of inner electrons at high frequencies $\omega>0.8$, leading to a non-monotonous behavior of the ionization rate not seen in the exact result.

\begin{figure*}[htbp]%
   \includegraphics[width=\textwidth]{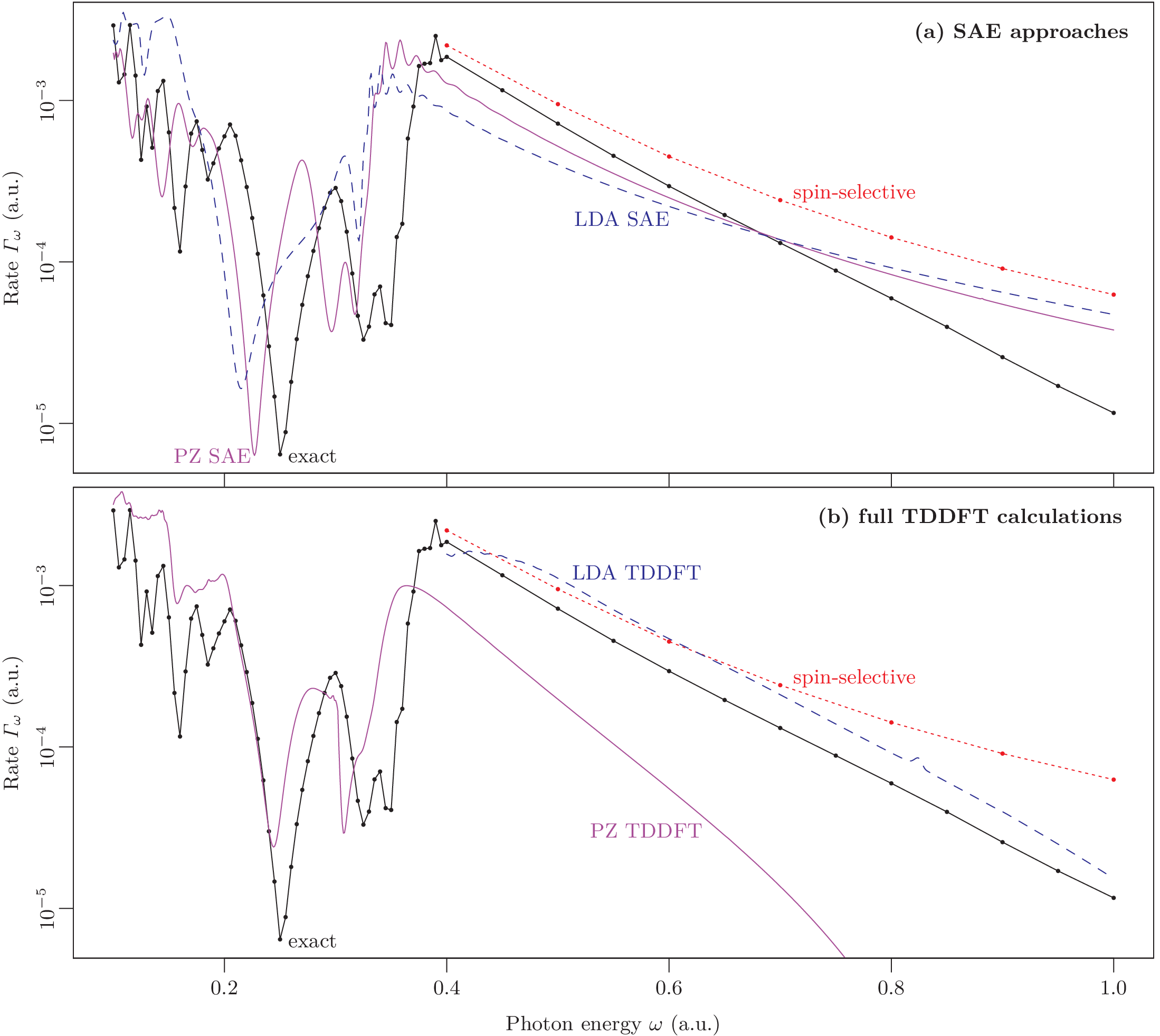}

   \caption{(Color online) Ionization rate $\itGamma_\omega$ obtained by SAE approaches (a) and by full TDDFT calculations (b). For reference, in both panels the result from the exact TDSE (labeled ``exact'', drawn solid with bullets) and the ``spin-selective'' TDSE calculation  (dotted with bullets)  in the single-photon regime $\omega>0.4$ where it is different from  the full  calculation (see Fig.~\ref{fig-rates-coupling}) are included.  
The SAE results in (a) are labeled ``LDA SAE'' (i.e., SAE with frozen LDA groundstate KS potential, dashed) and ``PZ SAE'' (i.e., SAE with frozen LDA PZ-SIC groundstate KS potential, solid). The full TDDFT results in (b) are labeled  ``LDA TDDFT'' (dashed) and ``PZ TDDFT'' (i.e., LDA with PZ-SIC, solid).  LDA TDDFT results are omitted in the multiphoton regime $\omega<0.4$,  PZ TDDFT in the high-frequency regime $\omega>0.8$, because a single-ionization rate cannot be determined in these cases.
}%
   \label{fig-rates-all}%
\end{figure*}%

\subsubsection{SAE results}\label{sec-sae-rates}
Ionization rates from SAE calculations in which frozen KS potentials were used are shown in Fig.~\ref{fig-rates-all}a.  In the low-frequency regime $\omega<0.2$ both SAE ionization rates change rapidly with the frequency, as the exact result does  due to the many avoided crossings discussed in Sec.~\ref{sec-floquet-results}. To the right of the two-photon ionization peak a sharp minimum at $\omega_\mathrm{min}\in[0.2, 0.26]$ shows up in all results. In the remaining  multiphoton ionization part up to $\omega=E_\mathrm{ip}\simeq 0.4$ one or two peaks are visible,  corresponding to one-photon excited-state-assisted ionization. 

The LDA SAE approach yields too few peaks and an incorrect  curvature around  $\omega\simeq 0.25$. Furthermore, the excited-state-assisted ionization peak around $\omega\simeq0.31$ is blue-shifted while the ionization threshold is red-shifted. This can be partially explained by the wrong asymptotic behavior of the KS potential originating from self-interaction in pure LDA. The SAE approximation in the PZ-corrected case leads to the correct number of peaks. If each of the $n$-photon-peaks is shifted to the right  by $\itDelta\omega\simeq\tfrac{0.04}{n}$ one finds a striking agreement with the exact TDSE result with respect to peak positions, widths, and heights. This improvement over pure LDA is due to the asymptotically correct KS potentials in the case of PZ SIC. In fact, three-dimensional DFT calculations applying the PZ SIC are often useful to quantitatively reproduce experimental values for ionization thresholds and excitation energies. Hence, the required shift $\itDelta\omega$ may be due to the one-dimensionality of the Li model system considered in this work.

In the single-photon ionization regime $\omega>E_\mathrm{ip}$ a monotonic decrease of the  ionization rate, starting from its maximum value for $\omega\simeq E_\mathrm{ip}$, is observed. However,  a convex curvature as for the spin-selective laser {\em gedankenexperiment} arises.  Hence, all approaches neglecting inner electron dynamics completely (pure LDA SAE and PZ-corrected SAE) or partially (spin-selective laser {\em gedankenexperiment}) yield a convex curvature in the logarithmic plot, presumably because the anti-screening effect is neglected.

\subsubsection{Full TDDFT results}\label{sec-full-rates}
Ionization rates from the full TDDFT calculations are shown in Fig.~\ref{fig-rates-all}b.
As mentioned above, plain LDA TDDFT does not allow to extract a rate in the multiphoton ionization regime at all. Compared to the SAE results, the  PZ TDDFT rate in the multiphoton ionization regime appears to be calculated with less spectral resolution. This is expected because ``unfrozen'' KS potentials do not support stationary energy levels that could aid ionization via resonant excitations. 

In the single-photon regime the full TDSE and the LDA TDDFT calculations yield an exponential decrease of the ionization rate for increasing photon energies. 
The corresponding slopes approximately equal each other. However, the rate predicted by LDA TDDFT is too high. A possible explanation for the overestimated rate is the following: the {\em down}shift of the KS energy during ionization in LDA without SIC (see Sec.~\ref{sec-sic-for-reso} below) leads to an {\em increased} ionization probability in the single-photon regime (because the ionization probability drops with increasing excess energy $\omega-E_\mathrm{ip}$). Hence, one may view the LDA rate as rather being blue-shifted than up-shifted.

Surprisingly, the PZ-corrected TDDFT calculations yield the worst rate, which is too small at the threshold, decreases too rapidly with increasing $\omega$, and has a concave curvature. The reason for this wrong behavior is discussed in Sec.~\ref{problem-pz-sic-single-photon} below.

Concluding this subsection, we can state that none of the considered approximations is able to yield correct ionization rates over the frequency interval $[0.1, 1.0]$, the best performing being the LDA PZ-SIC SAE approximation in the multiphoton and the pure LDA TDDFT in the single-photon ionization regime.

\subsection{Importance of  SIC for resonances in the multiphoton regime}\label{sec-sic-for-reso}
\begin{figure}[htbp]%
   \includegraphics[width=\columnwidth]{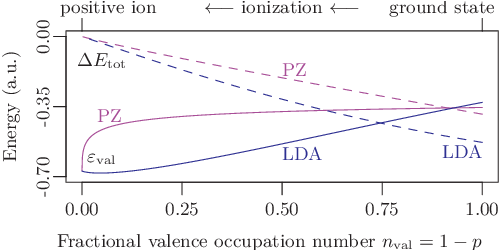}%
   \caption{(Color online) Orbital energy of the ``valence'' KS orbital $\varepsilon_\mathrm{val}$ (solid) and the total energy difference $\itDelta E_\mathrm{tot}$ (dashed) vs the fractional valence occupation number $n_\mathrm{val}=1-p$, where $p$ is the ionization probability, for pure LDA and for LDA with PZ SIC.}%
   \label{fig-occupation}%
\end{figure}

The PZ-SIC leads to ``better''  KS energies of the populated  levels in the sense of being closer to the respective ionization energies. The energies of the unpopulated levels in the groundstate KS potential also benefit from the SIC, leading to ``better'' excitation energies. Both are important ingredients for a correct ionization rate in the multiphoton regime. Moreover, in a time-dependent calculation SIC also helps taking into account the discontinuity of the KS potential  at integer orbital occupation numbers (this is the so-called ``derivative discontinuity'' in the XC energy, cf.~\cite{derDis}).   The ionization potential  must not depend on the ionization probability $p(t)$. Hence, the KS energy of the orbital from which predominantly ionization occurs should be independent of $p(t)\in[0, 1)$. Only when $p(t)=1$ is reached, the KS potential, and thus the orbital energy, should change discontinuously. The single-particle density in the vicinity of the nucleus decreases as $p(t)$ increases. As a result, the repulsive Hartree potential is weakened so that all orbital energies would shift to lower values if the XC potential did not counteract. In fact, pure LDA does not counteract properly so that the KS level energies vary continuously with $p(t)$. PZ SIC implements the derivative discontinuity, at least approximately. This is illustrated in Fig.~\ref{fig-occupation} for a valence KS orbital occupation $n_\mathrm{val}\in[0, 1]$ for pure LDA and PZ SIC applied to LDA. By considering fractional occupations in the stationary groundstate calculations  we are mimicking adiabatically evolving ionization, i.e., $1-n_\mathrm{val}=p$ with $p$ the instantaneous ionization probability.

For the neutral atom ground state $n_\mathrm{val}=1$ the PZ SIC lowers the valence orbital energy compared to the uncorrected LDA case,%
\begin{equation}%
 -0.355=\varepsilon_\mathrm{val}^{(\mathrm{PZ})}(1)<\varepsilon_\mathrm{val}^{(\mathrm{LDA})}(1)=-0.329.%
\end{equation}%
In the Li$^+$-limit $n_\mathrm{val}=0$  both approaches almost agree,%
\begin{equation}%
 -0.671=\varepsilon_\mathrm{val}^{(\mathrm{PZ})}(0)\simeq\varepsilon_\mathrm{val}^{(\mathrm{LDA})}(0)=-0.672.%
\end{equation}%
In pure LDA the orbital energy shifts almost linearly during ionization down to small values of $n_\mathrm{val}$. With PZ SIC  the orbital energy is shifted much less during ionization as long as $n_\mathrm{val}>0.2$. In the region $n_\mathrm{val}\in[0, 0.2]$ the PZ-SIC orbital energy describes a ``smoothed jump''  down to the Li$^+$-value. Hence, we find that the  PZ SIC smoothes the step-function-like behavior the unknown, exact SIC would yield.  The improvement over pure LDA regarding the constancy of the orbital energy of the ``ionizing KS level'' is essential.

\subsection{Problem of LDA PZ-SIC in single-photon ionization} \label{problem-pz-sic-single-photon}
With all the benefits from PZ SIC concerning KS level energies and the asymptotic behavior of the KS potential, it is an obvious question why PZ-corrected LDA fails so badly  in the single-photon ionization regime. The {\em gedankenexperiment} in Sec.~\ref{sec-gedankenexperiment} indicates that dynamical coupling effects between electrons such as anti-screening become increasingly important as the photon energy $\omega$ rises. In TDDFT, the coupling between KS orbitals is mediated by the Hartree part and the XC part of the KS potential. With PZ SIC  the orbital-dependent KS potential reads
\begin{eqnarray}
 v_i(x) &=& v(x) + v^\mathrm{(LDA)}[n_{\sigma_i}](x) + v^\mathrm{(H)}[n](x) \nonumber\\*
  &&- v^\mathrm{(LDA)}[n_i](x) - v^\mathrm{(H)}[n_i](x) \label{eq-pz-kohn-sham-potential}.
\end{eqnarray}
The Hartree potential is a linear functional of the total electron density
so that
\begin{eqnarray}
 v_i(x) &=& v(x) + v^\mathrm{(LDA)}[n_{\sigma_i}](x)\nonumber\\* 
&& - v^\mathrm{(LDA)}[n_i](x) + v^\mathrm{(H)}[n-n_i](x).
\end{eqnarray}
The effective Hartree term  $v^\mathrm{(H)}[n-n_i](x)$ for the  valence KS orbital after SIC  is solely determined by  the core KS orbital density. As a result, anti-screening is stronger than without SIC of the Hartree potential. The SIC of the LDA-part acts in the opposite direction. However, the SIC to the LDA XC potential is not  exact, so that a net overestimated anti-screening may remain,   leading to a lower ionization rate at higher photon energies.  Moreover, the dependence of the PZ TDDFT ionization rate as a function of the laser frequency is wrong in Fig.~\ref{fig-rates-all}b, pointing to a deficiency in the {\em dynamics} of the XC potential (note that the PZ SAE rate bends in the {\em opposite} direction). 

The fact that pure LDA leads at least to the correct slope of the ionization rate in the  single-photon ionization regime is thus likely due to a lucky cancellation of errors, i.e., the suppressed anti-screening is compensated by the error in the LDA XC potential   at all frequencies $\omega\in [0.5,1]$.

\section{Conclusion and outlook} \label{sec-concl}
In this work, we benchmarked various density functional-based approximate approaches to laser ionization with a numerically exactly
solvable three-electron model atom.

In the apparently simple photoeffect regime where only one photon is required for ionization, a surprisingly pronounced dependence 
of the ionization dynamics on the correct treatment of the inner electrons is found. These inner electrons are usually assumed to be passive, justifying frozen-core, single-active-electron, and pseudopotential approaches. However, because of the opposite timescales of inner and outer electrons with respect to the laser period {\em anti-screening} of the nuclear charge by the inner electrons occurs, which is ignored in such approaches (but could be modeled by a dynamical polarization potential). For instance, frozen-core orbitals lead to ionization rates too high, with an erroneously curved slope of the ionization rate as a function of the laser frequency. Unfortunately, the more advanced Perdew-Zunger self-interaction-corrected local density approximation fails as well at high frequencies because of an overemphasized anti-screening.

In order to correctly describe ionization in the multiphoton domain, energy levels as well as dipole transition 
probabilities must be reproduced by the simulation method, which is very demanding for pseudopotential and single-active-electron approaches. Moreover, the energy levels should not change as 
excited states get populated because this would move the system out of resonance. On the other hand, it is known that
once the population is inverted, the density is the ground state density of a ``new,'' discontinuously changing Kohn-Sham  potential \cite{PhysRevLett.89.023002-Maitra-Burke,tddft-rabi}. 
 Only proper
self-interaction free Kohn-Sham potentials may capture such multiphoton ionization effects involving resonances. 
We found that the Perdew-Zunger self-interaction-corrected local density approximation performs well for the lithium model atom in this respect, at least
qualitatively.

In future work, it is worth to compare our exact numerical model-Li results with Kohn-Sham results using more advanced exchange-correlation functionals than we did in the current work \cite{RevModPhys.80.3}. There might well be exchange-correlation potentials ``out there'' that perform well in both the multiphoton and single-photon ionization regime.

\section*{Acknowledgment}
This work was supported by the SFB 652 of the German Science Foundation (DFG). Discussions with V.~Kapoor are gratefully acknowledged.


\appendix

\section{Numerical solution of the three-particle TDSE}\label{sec-numerical-tdse}

\begin{figure}[htbp]%
   \includegraphics[width=\columnwidth]{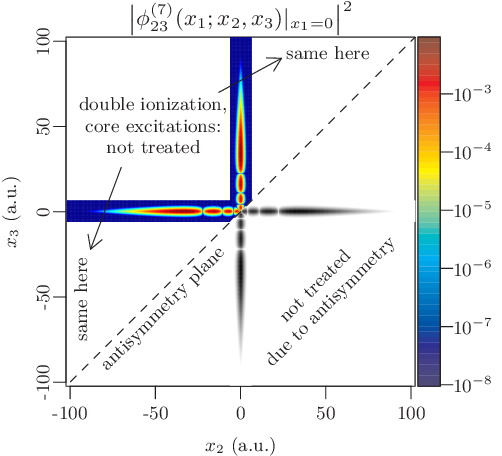}%
   \caption{(Color online) Spin-resolved probability density $|\phi_{23}^{(7)}(x_1, x_2, x_3)|^2$ of the seventh excited state cut at $x_1=0$. The adapted grid omits those regions with a white background.}%
   \label{fig-plot3d}%
\end{figure}%

In this Appendix, numerical details concerning the solution of the TDSE \eqref{tdse-like-equation} are given. The full solution of the three-particle TDSE for processes involving ionization is numerically demanding even if the spatial degrees of freedom are reduced to one per particle.  For that reason a Cartesian TDSE solver in three dimensions (i.e., one per particle) was implemented on a graphics processing unit (GPU) using the NVIDIA$^\text{\textregistered}$ CUDA$^\text{\texttrademark}$~\cite{cuda} platform. The solver is highly optimized for the purpose of this paper. A speedup of three orders of magnitude over a single-core CPU implementation covering the full 3D Cartesian grid is achieved on a desktop computer featuring an NVIDIA$^\text{\textregistered}$ GeForce$^\text{\textregistered}$ GTX 580 GPU. The speedup originates equally from an efficient implementation utilizing the high single-precision performance of the GPU and the adjustment of the simulation grid. ``Mixed precision'' techniques allow to obtain results in double precision although most of the numerical effort consists of single-precision floating point operations.

The kinetic energy operator is discretized using the implicit Numerov expression, which is accurate up to fourth order in the spatial gridspacing $\itDelta x$. For the propagation in time we employ the unitary Crank-Nicolson method consisting of an (explicit) forward and an (implicit) backward step, accurate up to second order in the timestep ${\itDelta t}$. The explicitly  time-dependent Hamiltonian $H(t)$ is evaluated at midpoints $\tau=t+\tfrac{\itDelta t}{2}$. 

The linear equations to propagate the discretized wavefunction in time for one timestep can easily be arranged to require the solution of no more than a single implicit equation with a $3\times3\times3$ stencil. For a sufficiently small timestep $\itDelta t$ the corresponding coefficient matrix is diagonal-dominant. Hence, the Jacobi method can be used to determine its solution. Faster convergence of long-wavelength errors is achieved by applying a multigrid scheme. Finally, a high throughput of floating point operations is obtained by using mixed precision techniques and different data caching stages combined with massive parallelization.

The numerical grid was optimized, exploiting what is known about the probability density dynamics during long-wavelength, single ionization. Recall that the laser is tuned such that solely the outer electron is removed from the atom. As a consequence, the single spin-down inner electron is tightly bound to the core at all times. This means that $|\phi_{23}(t; x_1, x_2, x_3)|^2$ will only yield non-vanishing probabilities for small $|x_1|$. Hence, one may choose a small box size in the  $x_1$-direction. In contrast to that, the inner-outer spin component in the $x_2$ and $x_3$-directions may be spatially extended. However, if both $|x_2|$ \emph{and} $|x_3|$ are large, $|\phi_{23}(t; x_1, x_2, x_3)|^2$ must be negligible  for the allowed laser parameters, i.e., those  that do {\em  not} lead to double  (or triple) ionization.
 Consequently, the grid regions where $|x_2|$ {\em and} $|x_3|$ are large are omitted.
Finally, by making use of the antisymmetry $\phi_{23}(t; x_1, x_2, x_3)=-\phi_{23}(t; x_1, x_3, x_2)$ only the region $x_2\leq x_3$ needs to be considered numerically.

Denoting the width of an ionization channel in grid points by $N_\mathrm{small}$ and its total length by $N_\mathrm{large}$, the total number of gridpoints compared to an $N_\mathrm{small}\times N_\mathrm{large}\times N_\mathrm{large}$ cuboid is reduced by a factor of $2 N_\mathrm{large}/N_\mathrm{small}$. The results presented in this work have been checked to be  converged for $N_\mathrm{small}=64$ and $N_\mathrm{large}=1024$ (for $\itDelta x=0.2$ and $\itDelta t= \itDelta x/8$), which corresponds to a speedup of $32$ caused by the grid adjustment alone. As an example,  the probability density $|\phi_{23}^{(7)}(x_1=0, x_2, x_3)|^2$ of the seventh excited state on the optimized grid is shown in  Fig.~\ref{fig-plot3d}.

\end{document}